\documentclass[]{spie}  

\usepackage{amsmath,amsfonts,amssymb}
\usepackage{graphicx}
\usepackage[colorlinks=true, allcolors=blue]{hyperref}
\usepackage{xcolor}
\usepackage[normalem]{ulem}

\title{CCAT-prime: RFSoC Based Readout for Frequency Multiplexed Kinetic Inductance Detectors}

\author[a]{Adrian K. Sinclair}
\author[b]{Ryan C. Stephenson}
\author[c]{Cody A. Roberson}
\author[c]{Eric L. Weeks}
\author[a]{James Burgoyne}
\author[d]{Anthony I. Huber}
\author[b,c]{Philip M. Mauskopf}
\author[a,d,e]{Scott C. Chapman}
\author[g]{Jason E. Austermann}
\author[f]{Steve K. Choi}
\author[f]{Cody J. Duell}
\author[h]{Michel Fich}
\author[c]{Christopher E. Groppi}
\author[f]{Zachary Huber}
\author[f]{Michael D. Niemack}
\author[i]{Thomas Nikola}
\author[i]{Kayla M. Rossi}
\author[c]{Adhitya Sriram}
\author[j]{Gordon J. Stacey}
\author[f]{Erik Szakiel}
\author[a]{Joel Tsuchitori}
\author[f]{Eve M. Vavagiakis}
\author[g]{Jordan D. Wheeler}
\author{the CCAT-prime collaboration}

\affil[a]{Dept. of Physics and Astronomy, University of British Columbia, Vancouver, BC, Canada} 
\affil[b]{Dept. of Physics, Arizona State University, Tempe, AZ, USA}
\affil[c]{School of Earth and Space Exploration, Arizona State University, Tempe, AZ, USA} 
\affil[d]{Dept. of Physics and Astronomy, University of Victoria, Victoria, BC, Canada }
\affil[e]{Herzberg Astronomy and Astrophysics Research Centre, National Research Council Canada, Victoria, BC, Canada}
\affil[f]{Dept. of Physics, Cornell University, Ithaca, NY, USA }
\affil[g]{National Institute of Standards and Technology, Boulder, CO, USA }
\affil[h]{Waterloo Centre for Astrophysics, University of Waterloo, Waterloo, ON, Canada}
\affil[i]{Cornell Center for Astrophysics and Planetary Sciences, Cornell University, Ithaca, NY, USA}
\affil[j]{Dept. of Astronomy, Cornell University, Ithaca, NY, USA}

\authorinfo{Further author information: (Send correspondence to A.K.S.)\\A.K.S.: E-mail: adriansinclair@phas.ubc.ca}

\begin{document} 
\maketitle

\begin{abstract}
The Prime-Cam instrument on the Fred Young Submillimeter Telescope (FYST) is expected to be the largest deployment of millimeter and submillimeter sensitive kinetic inductance detectors to date. To read out these arrays efficiently, a microwave frequency multiplexed readout has been designed to run on the Xilinx Radio Frequency System on a Chip (RFSoC). The RFSoC has dramatically improved every category of size, weight, power, cost, and bandwidth over the previous generation readout systems. We describe a baseline firmware design which can read out four independent RF networks each with 500 MHz of bandwidth and 1000 detectors for $\sim$30 W. The overall readout architecture is a combination of hardware, gateware/firmware, software, and network design. The requirements of the readout are driven by the 850 GHz instrument module of the 7-module Prime-Cam instrument. These requirements along with other constraints which have led to critical design choices are highlighted. Preliminary measurements of the system phase noise and dynamic range are presented.


\end{abstract}


\section{Introduction}
The far-infrared sky, loosely defined from $\sim$100 GHz to a few THz, has been a challenging band of the electromagnetic spectrum to observe. Atmospheric absorption and emission require strategic band selection and extraordinarily dry and high altitude observation sites. In part due to these challenges the far-infrared (far-IR) sky still has many motivating and clear science goals\cite{CCATSciForcast2021}, a few of these are highlighted below. The role of turbulence and magnetic fields can be studied through the polarized far-IR emission of interstellar dust grains. The variability in protostars which informs stellar accretion and planet formation models can be studied from mid to far-IR. Galaxy formation and star formation history can be tracked through redshifted lines such as CO, OIII, and CII, which fall into the same band. High redshift proto-clusters of galaxies can be probed via the kinematic Sunyaev-Zel'dovich effect at hundreds of gigahertz. On the largest scales, the cosmic microwave background requires well calibrated sub-millimeter observations of polarized dust emission to remove its contaminating foregrounds. These are just a few of the motivating science topics for the Fred Young Submillimeter Telescope and first light instruments described next.

Seated high atop Cerro Chajnantor in the Atacama desert of Chile, the CCAT-prime observatory will include the Prime-Cam instrument on the Fred Young Submillimeter Telescope (FYST) and will map the sky with imaging polarimeters\cite{Vavagiakis2018,Choi2021, Scott2022} and spectrometers\cite{Cothard2020, Nikola2022}. The telescope's wide field of view and large optical throughput\cite{Niemack2016, Parshley2018} enables high detector count focal planes. Mod-Cam\cite{Duell2020} is a single instrument module receiver that is a precursor and testbed for Prime-Cam and is planned to achieve first light in 2024. The enabling detector technology is the Lumped Element Kinetic Inductance Detector (LEKID) which, due to its natural multiplexing, low fabrication cost, and low cryogenic readout complexity, can be scaled to the $>$ 100,000 detectors required for Prime-Cam. A second enabling technology is the Xilinx Radio Frequency System-on-Chip (RFSoC) which has dramatically improved every category of size, weight, power, cost, and bandwidth over the previous generation KID readout systems. 

Kinetic Inductance Detectors (KIDs) were developed nearly twenty years ago \cite{Day2003} to enable ultra-wideband astronomical imaging at low fabrication cost with high multiplexing factors into the thousands. Since then, single photon sensitivity with modest energy resolution has been achieved from X-rays to the near-infrared \cite{Mazin2006,Szypryt2017,deVisser2021}. At longer wavelengths in the millimeter and submillimeter, photon noise limited performance has been shown in linear power detection mode \cite{Yates2011, Hubmayr2014, Mauskopf2014, Dober2016, Flanigan2016}. There are recent encouraging results with sub-gap absorption KIDs\cite{Levy-Bertrand2021}, DC biasing \cite{Zhao2020}, and bi-layers \cite{Aja2021}, opening up even longer wavelength detection typical of CMB studies. KIDs have been antenna coupled \cite{Day2006, Barry2018}, horn coupled \cite{McCarrick2014, Bryan2015, Austermann2018}, and packed into filled arrays\cite{Roesch2011}, each achieving high optical efficiency to allow dense focal plane packing. A shift from distributed to a lumped element kinetic inductance detector (LEKID) \cite{Doyle2008} design has simplified frequency placement and relaxed simulation requirements. The majority of the current and previously deployed mm/sub-mm imaging telescopes which utilize KIDs have adopted lumped element designs \cite{Swenson2012, Monfardini2011, Adam2018, Coppi2020, Paiella2020, TolTEC2020, Brien2018}. Prime-Cam will employ a lumped element design that is horn coupled capitalizing on the development experience of the BLAST-TNG and TolTEC LEKID arrays \cite{Austermann2018}.

Early KID readouts leveraged advancements in commercially available software defined radio technology \cite{MazinSDR2006} and previously developed radio astronomy spectrometers \cite{Yates2009, Klein2006}. Custom boards started to be developed in the early 2010s \cite{Bourrion2011, Bourrion2016,Rantwijk2016}. A reconfigurable open-architecture computing hardware (ROACH) based system, originally designed for radio astronomy, was repurposed by multiple groups \cite{Swenson2012, Duan2020, McHugh2012}, as well as the second generation of the board (ROACH2) a few years later \cite{Gordon2016, Strader2016,Fruitwala2020}. The open source firmware developed for the ROACH2 based readout of BLAST-TNG\cite{Gordon2016} has enjoyed use in the following experiments: OLIMPO\cite{Paiella2019}, MUSCAT\cite{Tapia2019}, SuperSpec\cite{Redford2021}, and TolTEC\cite{TolTEC2020}. The RFSoC by Xilinx arrived near the end of the decade and immediately generated interest in the community and spurred multiple parallel development efforts \cite{Smith2021,Baldwin2020, Stefanazzi2021, Bradley2021, Sinclair2020, Frisch2022}. This work presents an RFSoC based readout to be to be deployed on FYST for the Mod-Cam and Prime-Cam instruments and a future on-chip spectrometer.
\section{Requirements}
\label{sec:reqs}
The main driver of the readout system requirements are the detectors in five ways: Total number of detectors, number of detectors per RF network, bandwidth per network, added noise, and crosstalk. There are more stringent requirements for the frequency comb discussed in the gateware Sec. \ref{sec:freq_comb}. For the 280 GHz, 350 GHz, and EoR-Spec detector arrays the baseline bandwidth of $\sim$500 MHz with 600 resonators or less per RF network has been adopted. In order to adhere to thermal budget constraints, Prime-Cam's highest frequency instrument - the 850$~$GHz module\cite{Tony2022} is planning for two octaves of bandwidth (1 GHz) for each RF network.  The 850 GHz arrays push the bandwidth requirements from the detectors to the readout while previous KID arrays developed for BLAST-TNG\cite{Hubmayr2014} and TolTEC\cite{Austermann2018} have been constrained by the instantaneous bandwidth of the ROACH2 based readout\cite{Gordon2016}. The readout noise and crosstalk requirements are derived from detector parameters in the next two sections.  

\subsection{Readout Noise Requirement}
The readout system should not contribute significantly to the noise, which places a requirement on both the noise and dynamic range of the system. We use the method of Sipola et al.\cite{Sipola2019} to calculate an equivalent output noise temperature of the KID and compare this to other readout contributors such as the cryogenic low noise amplifier (LNA) and digital to analog converter (D/A). The detector equivalent output noise temperature is defined as,
\begin{equation}
    T = \frac{(R \times \textrm{NEP})^2}{kZ_0},
\label{eq:output_noise_temp}
\end{equation}
where $R$ is the voltage responsivity to power, NEP is the detector noise equivalent power, $k$ is Boltzmann's constant, $Z_0$ is the circuit's characteristic impedance of 50 $\Omega$. 
The voltage responsivity can be derived from the complex forward transmission of a resonator \cite{Mauskopf2018},
\begin{equation}
    S_{21} = 1 - \frac{Q_r}{Q_c}\frac{1}{1 + j2Q_r x},
\end{equation}
where $Q_r$ and $Q_c$ are the total and coupling quality factors respectively, and $x$ is the fractional frequency shift. The voltage at the output of the resonator would be the product of the voltage incident on the resonator with the transmission, $V_{out}=V_{in}S_{21}(x)$. Appendix \ref{sec:eq_det_noise} shows that the resulting voltage responsivity to power is then
\begin{equation}
    \frac{dV_{out}}{dP} = V_{in}\frac{dx}{dP}\Big( \frac{d Re\{ S_{21}\}}{dx}  + j\frac{d Im\{ S_{21}\}}{dx} \Big ),
\end{equation}
where $V_{in}$ is the incident voltage of the bias tone, and $dx/dP$ is the fractional frequency responsivity to optical power. 

The two main readout noise contributors considered here are the LNA and D/A. We compare the noise at a reference plane after the resonator array and before the LNA. This is convenient for the LNA as the measured noise temperature reported is referred to the input of the amplifiers. The amplifiers to be used in initial deployment have input noise temperatures of approximately 5 K across the band \cite{Hamdi2014}. The D/A noise however depends on the total attenuation to the reference plane. The quantization noise power can be estimated from the variance of the error first derived by Bennett\cite{Bennett1948} in 1948 as
\begin{equation}
    P_{digi} = \frac{\Delta V^2}{12 Z_0} = \frac{1}{12Z_0}\frac{V_{FS}^2}{2^{2N_{bits}}},
\end{equation}
where $V_{FS}$ is the full scale voltage of the digitizer, $N_{bits}$ are the effective number of bits. This can be converted into an equivalent noise temperature by dividing by the Nyquist bandwidth $f_s/2$, Boltzmann's constant $k$, and scaling by the total attenuation $A$ to the reference plane to get,
\begin{equation}
    T_{digi} = A\frac{P_{digi}}{k f_s/2}.
\end{equation}

\begin{figure}[!t]
\centering
\includegraphics[width=4in]{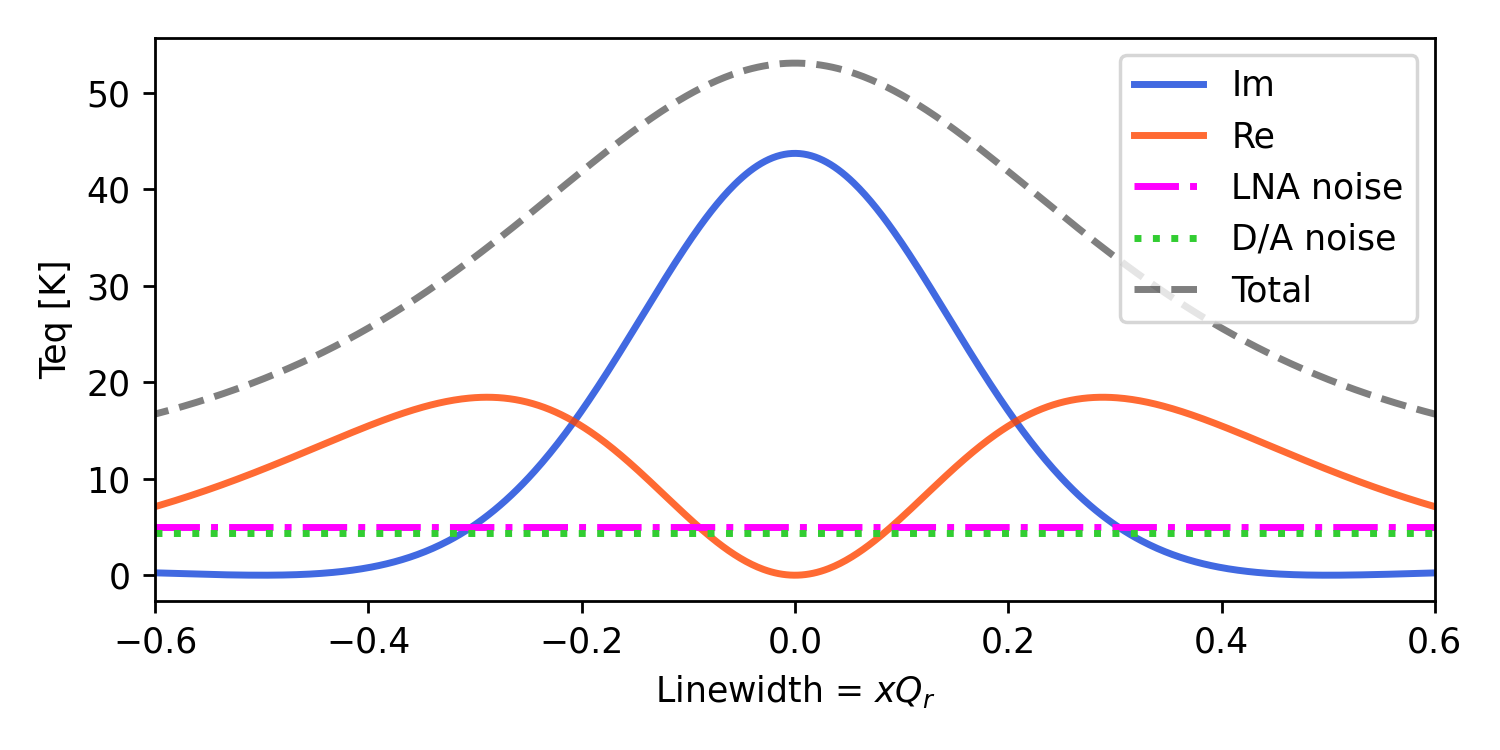}
\caption[Equivalent Noise Stackup]{The effective noise temperature of the KID in both quadratures, LNA and D/A noise, plotted against the shift from resonance in multiples of a linewidth. The dashed grey line represents the total noise summing D/A, LNA, Re, and Im.}
\label{fig:noise_stackup}
\end{figure}

Using the measured in-flight median values for the BLAST-TNG 850 GHz array as a stand in for the Prime-Cam 850GHz arrays, which are currently under development, we plot the noise temperature vs shift in linewidths in Fig. \ref{fig:noise_stackup}. The parameters used for Fig. \ref{fig:noise_stackup} are: total quality factor $Q_r=36000$, coupling quality factor $Q_c=58300$, center resonant frequency $f_0 = 800$ MHz, NEP = $10^{-16}$ $W/\sqrt{Hz}$, fractional frequency responsivity $dx/dP=5.5 \times 10^{-6}$ $x/pW$, $V_{FS}=0.7$ V, $N_{bits}=12$, attenuation from D/A to detectors A = 35 dB, tone power at detectors $P_{tone}=-90$ dBm, and a sampling frequency fs=512 MHz. Figure \ref{fig:noise_stackup} plots both components of the complex output noise temperature from Eq. \ref{eq:output_noise_temp} as $Re$ and $Im$. In typical operation the complex signal is converted such that the response is mainly in one quadrature: the output noise temperature of the detectors is $\sim$40K in the $Im$ quadrature on resonance. As the detector shifts away from its previous resonant frequency the output noise temperature decreases until falling to the level of the other contributors. LNA noise and D/A noise are approximately equal at the output of the detector arrays. 

\begin{figure}[!t]
\centering
\includegraphics[width=4in]{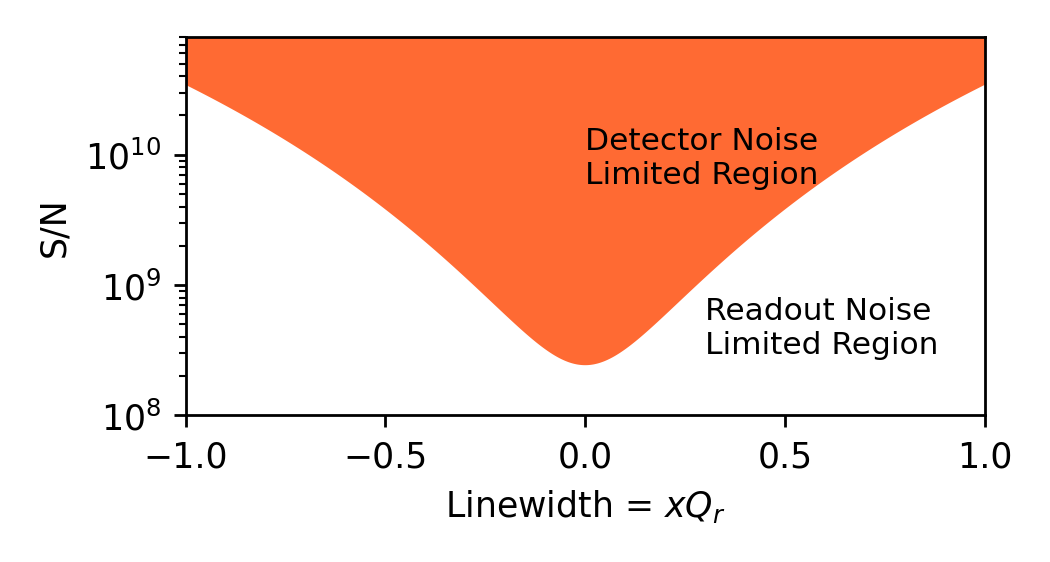}
\caption[S/N vs Fractional Frequency Shift]{Signal to noise ratio from Eq. \ref{eq:snr_det} measured at the output of the detector plotted against the shift from resonance in multiples of a linewidth. The solid orange section above the curve is considered the detector noise limited region and anything below or outside would be limited by the readout. For example if a readout system has a signal to noise of 100 dB ($10^{10}$) then it will be detector noise limited for shifts up to approximately $\sim$65\% of a linewidth.}
\label{fig:snr_vs_frac_freq}
\end{figure}

To find the required dynamic range for the readout we must find the maximum signal to noise measured at the output of the detectors. This is calculated as the ratio of the bias tone power scaled by the detector transmission to the output detector noise power,
\begin{equation}
    S/N = \frac{P_{tone}|S_{21}(x)|^2}{kT_{det}} = \frac{|S_{21}(x)|^2}{\Big (\frac{dx}{dP}\Big|\frac{dS_{21}(x)}{dx}\Big|NEP \Big )^2}.
\label{eq:snr_det}
\end{equation}
The tone power drops out of the equation and the signal to noise is determined solely by properties of the detector. Plotting Eq. \ref{eq:snr_det} as a function of linewidth with the same detector parameters from above produces the curve in Fig. \ref{fig:snr_vs_frac_freq}. The minimum signal to noise measured at the output of the detectors occurs on resonance and increases further from this point. The readout systems dynamic range must be greater than the curve defined by Eq. \ref{eq:snr_det} to not degrade the total signal to noise. Figure \ref{fig:snr_vs_frac_freq} shows that if the readout system has a measured dynamic range of 100 dB then detector noise limited performance is possible for shifts of up to $\sim 65 \%$ of a linewidth. If the shifts are greater than that however, the signal to noise will become readout noise limited.

\subsection{Channel Crosstalk}
\label{sec:xtalk_reqs}  
\begin{figure}[!t]
\centering
\includegraphics[width=4in]{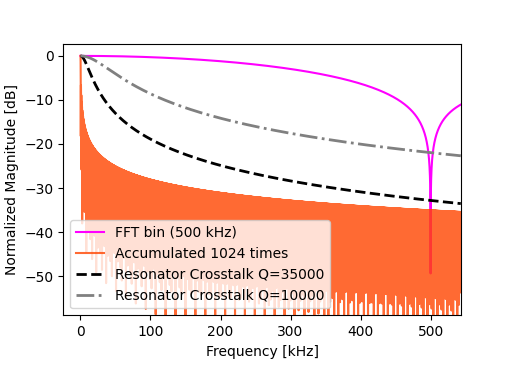}
\caption[Accumulated Spectrum with X-talk]{Accumulated spectrum with detector crosstalk for two different quality factors overlaid with the frequency response of the coarse and fine channelization (see Sec. \ref{sec:fine_channelization}).}
\label{fig:accum_spectrum_and_xtalk}
\end{figure}

Kinetic inductance detectors suffer from electrical crosstalk when their frequency responses overlap. This is minimized by a combination of accurate frequency placement, high quality factors, and post-fabrication editing \cite{Liu2017, shroyer2022}. We can calculate the overlap between two identical resonators which are frequency neighbors as,
\begin{equation}
    X_{talk} = \frac{|S_{21}^{n}(f_0)-1|^2}{|S_{21}(f_0)-1|^2} = \frac{1}{1 + 4 N^2},
\label{eq:xtalk}
\end{equation}
where $S_{21}^{n}$ is the forward transmission of the neighboring resonator evaluated at the resonant frequency $f_0$ of the other and normalized by the on resonance transmission, N is the number of linewidths ($N=Q_r x$) of separation between the adjacent resonators. This electrical crosstalk takes place at the detector arrays and sets a fundamental requirement for the digital signal processing to meet. Figure \ref{fig:accum_spectrum_and_xtalk} plots Eq. \ref{eq:xtalk} for two different quality factors against the frequency response of the coarse and fine channelization stages within the gateware (see Sec. \ref{sec:gateware}). The figure shows that a Fast Fourier Transform (FFT) based coarse channelization with a bin width of 500 KHz shown in magenta, does not meet the spec of both quality factors without an additional fine channelization stage. The frequency response of a fine channelization stage with 1024 accumulations is plotted in orange displaying sidelobes which fall below the intrinsic detector crosstalk for both quality factors. This suggests that for high rate applications such as pulse detection, side lobe minimizing windows or polyphase filterbanks would be necessary to keep channel leakage below the detector crosstalk levels.

\section{Baseline Design}
\subsection{Overall architecture}
\begin{figure}[!t]
\centering
\includegraphics[width=4in]{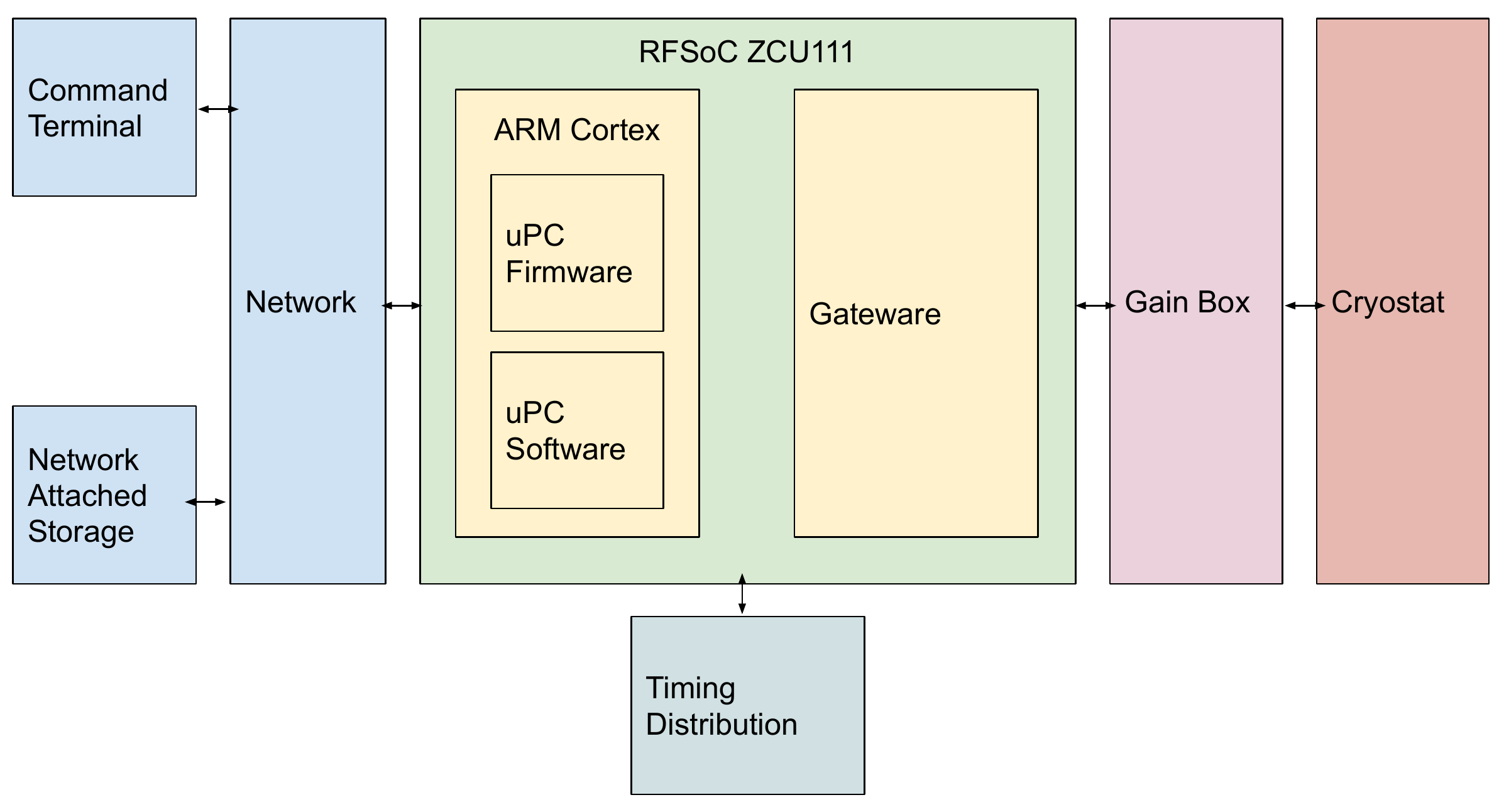}
\caption[Overall Architecture Block Diagram]{Block diagram of the overall readout architecture showing the main systems and interconnections between each.}
\label{fig:overall_arch}
\end{figure}

The baseline design will use the Xilinx RFSoC which contains reconfigurable logic fabric, an ARM microprocessor (uPC), and high speed digitizers all integrated into a single chip. The first generation development board housing this chip (ZCU111) is the chosen platform for CCAT-prime due to its cost and familiarity from previous efforts by the author\cite{Sinclair2020, Sinclair2021}. The digital signal processing gateware which accomplishes the stimulus and demodulation of the detectors will sit in the reconfigurable fabric. The ARM uPC will run a lightweight version of Linux (PetaLinux) as an operating system. The software which directly interfaces the gateware and performs various array calibration tasks will reside on this ARM core. Many independently controlled RFSoC ZCU111 development boards will be operating simultaneously in the deployed system, with all communications occurring through an Ethernet network. A precision and stable time source will be distributed to each RFSoC to ensure pointing and detector data synchronization. The overall architecture can be roughly split into the following sections: hardware, gateware, uPC firmware, uPC software and networking, and analysis. A block diagram illustrating these systems and their general interconnections is shown in Fig. \ref{fig:overall_arch}. 

\subsection{Hardware}
Most hardware components of the readout are commercially available but others require development including, an SMA breakout board for the ZCU111, a peripheral module sync board, and a rack-mountable enclosure. An optimized readout also depends on the cryogenic hardware such as the LNA, attenuators, and coaxial cables. The cryogenic attenuation to the arrays must reduce the room temperature thermal noise but not reduce the dynamic range of the tones produced by the D/A. The optimal cryogenic gain from the detectors to the A/D to prevent saturation should be balanced with available cooling power. See \citenum{Tony2022} for more information on this for Prime-Cam's 850 GHz instrument. 

The ZCU111 development board and an SMA breakout board will be housed in a 1U rack-mountable enclosure with Ethernet, power, and SMA ports. Cerro Chajnantor has half an atmosphere of pressure and the instrument cabin will not be pressurized. This requires the RFSoC system to dissipate at high altitude without overheating. A method which was used for the balloon-borne project BLAST-TNG was to conduction cool with copper heat blocks with water filled heat pipes dumping the heat to the gondola frame. In this case the heat sink would be plates on the bottom of the enclosures and the electronics rack itself. Each enclosure will have 10 SMA ports (x8 for A/D and D/A, x1 for 10MHz reference, x1 for sync input), two 1GbE ports (x1 commanding, x1 streaming UDP detector packets), and one power port.

To allow for analog sync inputs to the FPGA fabric on the ZCU111, a small board which connects to the Peripheral Modules (PMOD) connector has been designed. The analog sync could come from a pulse-per-second of a GPS disciplined oscillator or halfwave plate or pointing sensor trigger. This signal would be digitized and incorporated with other timing signals such as free running counters and packets counters into the final data UDP packet.

Each RFSoC ZCU111 development board comes with the XM500 SMA breakout board, this has a variety of different connections in an effort to allow various use cases. However we would like to take the differential signals from the RF data converter pins and pass all of them through low frequency baluns (10 - 3000 MHz) to edge mounted SMA connectors. A prototype PCB of the new design is currently being tested. 

\subsection{Gateware}
\label{sec:gateware}
\begin{figure}[!t]
\centering
\includegraphics[width=4in]{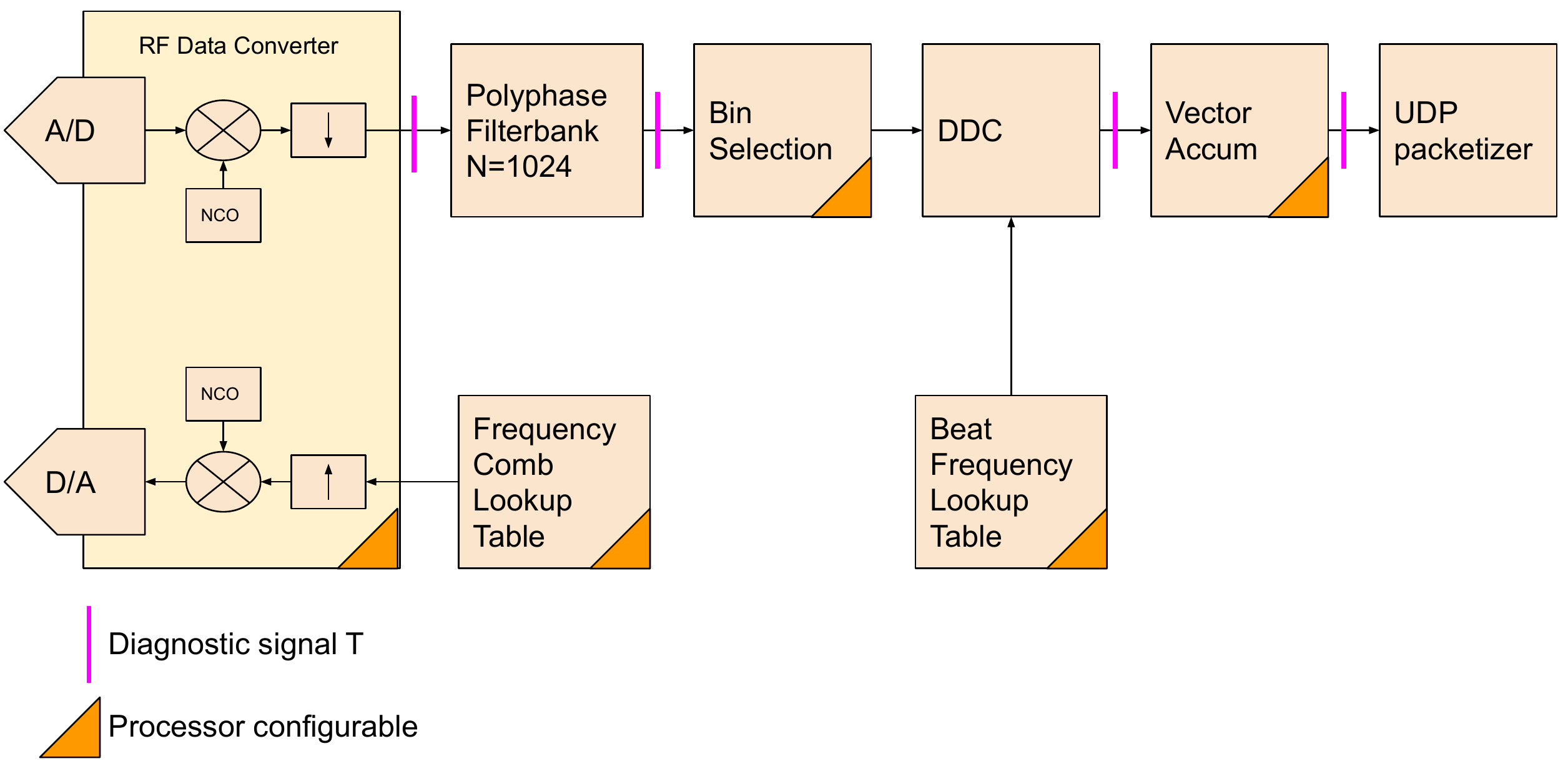}
\caption[Gateware Block Diagram]{Block diagram of the digital signal processing gateware and connection to RF data converters implemented on the RFSoC programmable logic.}
\label{fig:gateware_block_design}
\end{figure}

The gateware (FPGA firmware) is responsible for generating the detector array stimulus and producing demuxed, averaged, and downsampled detector data. The heritage for the design leans heavily on the design for BLAST-TNG \cite{Gordon2016}, and the pioneering designs of MAKO\cite{Swenson2012}, ARCONS \cite{McHugh2012}, and MUSIC \cite{Duan2020}. The digital signal processing chain consists of a lookup table which stores the frequency comb, two stage channelization accomplished with a polyphase-filterbank followed by a  digital downconversion and accumulator. This section is split up between the different modules roughly outlined by the block diagram shown in Fig. \ref{fig:gateware_block_design}.

\subsubsection{Frequency Comb Lookup Table}
\label{sec:freq_comb}
As the main stimulus for the detectors this part of the gateware design is closely linked to the array properties and their requirements. The frequency comb must be able to produce tones with high resolution that is significantly less than the resonator FWHM anywhere within the Nyquist bandwidth. Additional desirable features are quick re-writing of the waveform and to uniquely assign phase and amplitudes to each tone. To maximize the available D/A dynamic range the waveform crest factor and normalization method should be carefully constructed to not lose more than a typical OFDM peak-to-average power ratio of 12 dB.

The planned method of frequency comb generation is to utilize the programmable logic (PL) interfacing DDR4 memory. This is a large memory bank which operates at four times the PL fabric clock rate thus delivering four simultaneous samples. The initial reason for this choice is that in order to avoid spurious signals in the waveform we enforce phase continuity on the wrap around (meaning when the memory address to read goes from the maximum address to the minimum). A discontinuity in the time domain increases the spurious floor in the frequency domain. This can limit the dynamic range of the system and lower the mux factor. As an example, the baseline design has an effective Nyquist bandwidth of 512 MHz, and to generate a phase continuous waveform with a tone resolution of $\Delta f \sim 500$Hz it would require a lookup table size of:
\begin{equation}
    LUT = f_s/\Delta f \approx 1 \ \ \textrm{Million.}
\end{equation}
This is reasonable to place within the FPGA fabric for a single DSP chain but will not be when implementing four DSP chains. Therefore this must be placed in an external memory such as the PL connected DDR4.

\subsubsection{RF Data Converter Configuration - Digital Mixers}
The RF data converter in the Zynq UltraScale+ XCZU28DR-2FFVG1517E RFSoC \cite{UG1287} has a compile time configuration which allows for the use of internal digital mixers with a numerically controlled oscillator (NCO) as well as decimation and interpolation filters. This is shown as the RF Data Converter block at the front of the diagram in Fig. \ref{fig:gateware_block_design}. The high speed sampling provide direct RF digitization without the need for analog mixers, eliminating the need for an intermediate-frequency (IF) system with analog mixers.

\begin{figure}[!t]
\centering
\includegraphics[width=4in]{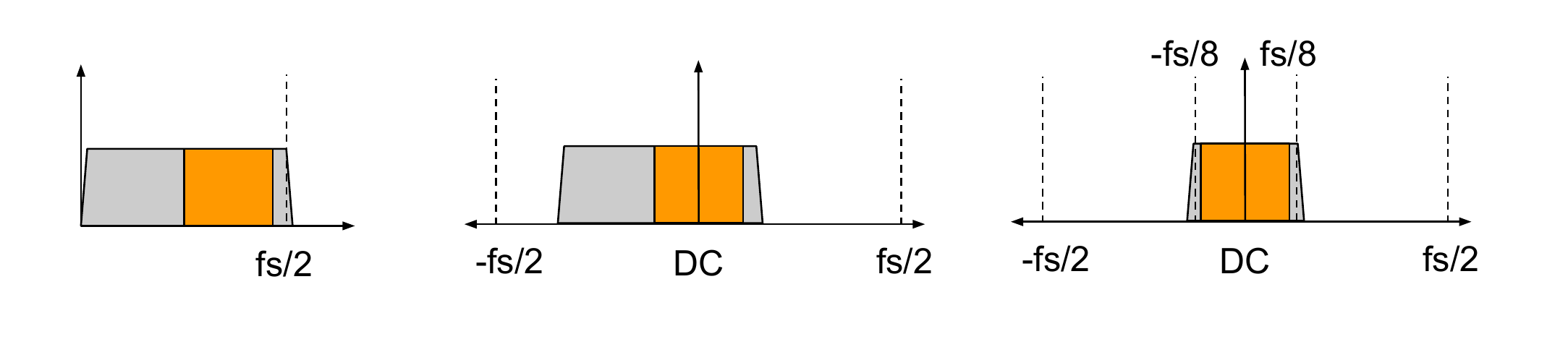}
\caption[RF data converter]{Three steps accomplished by the RF data converter for the A/D starting at the left and for the D/A starting from the right. The first spectrum is showing the Nyquist bandwidth from 0-$f_s/2$ with some band of interest in orange. The second spectrum is post digital down-conversion to some frequency at the center of the band of interest accomplished with a numerically controlled oscillator (NCO). The third is the filtered and downsampled spectrum. In the reverse order, interpolation and up-conversion to the D/A.}
\label{fig:rf_data_conv}
\end{figure}

To understand how the RF data converters will be used in the baseline design, the three processing steps are shown in Fig. \ref{fig:rf_data_conv}. In the first step starting on the left is the directly digitized Nyquist bandwidth of $f_s/2$. The orange block represents the band of interest that contains the resonators. Directly after digitization the signals get passed to the digital mixer with NCO. This NCO is tuned to match the center frequency of the band of interest thereby mixing the band center to DC, this is shown in the middle spectrum of the same figure. The mixer outputs a complex signal which is a combination of in-phase and quadrature complex signals which are then filtered and decimated, shown in the last figure as a factor of 2. This step is critical in that it allows us to pass only the required bandwidth through the FPGA gateware reducing the number of parallel paths and resource utilization. The same three steps are done in reverse for the generation of the frequency comb from the D/A.

\subsubsection{Coarse Channelization}
The first step in de-multiplexing is to perform an N-point Fast-Fourier Transform (FFT) which splits the input timestream into N channels of bandwidth $f_{samp}/N$, Where $f_{samp}$ is the sampling rate post RF data converter. Each complex output from the FFT is considered a bin. The frequency response of a single FFT bin takes a sinx/x form with $x=\pi n$ where n is the ratio of sample rate and input frequency. This frequency response assures that the adjacent bin centers are zero at the cost of having a loss of signal at the bin edge and high side lobes. To prevent this loss and substantially lower the side band leakage many groups have employed the weighted overlap and add (WOLA) polyphase filter-bank (PFB). While the WOLA PFB recovers the bin edge attenuation and provides superior side lobe performance, its typical implementation requires significantly more resources than simple window functions. Earlier in Sec. \ref{sec:xtalk_reqs} we showed that for lower bandwidth detector readout ( $\sim$ 500 Hz) that employs a fine channelization stage with many accumulations the frequency response of the coarse stage becomes insignificant. Future work will compare the PFB vs FFT baseline.  

\subsubsection{Bin Selection}
\begin{figure}[!t]
\centering
\includegraphics[width=3in]{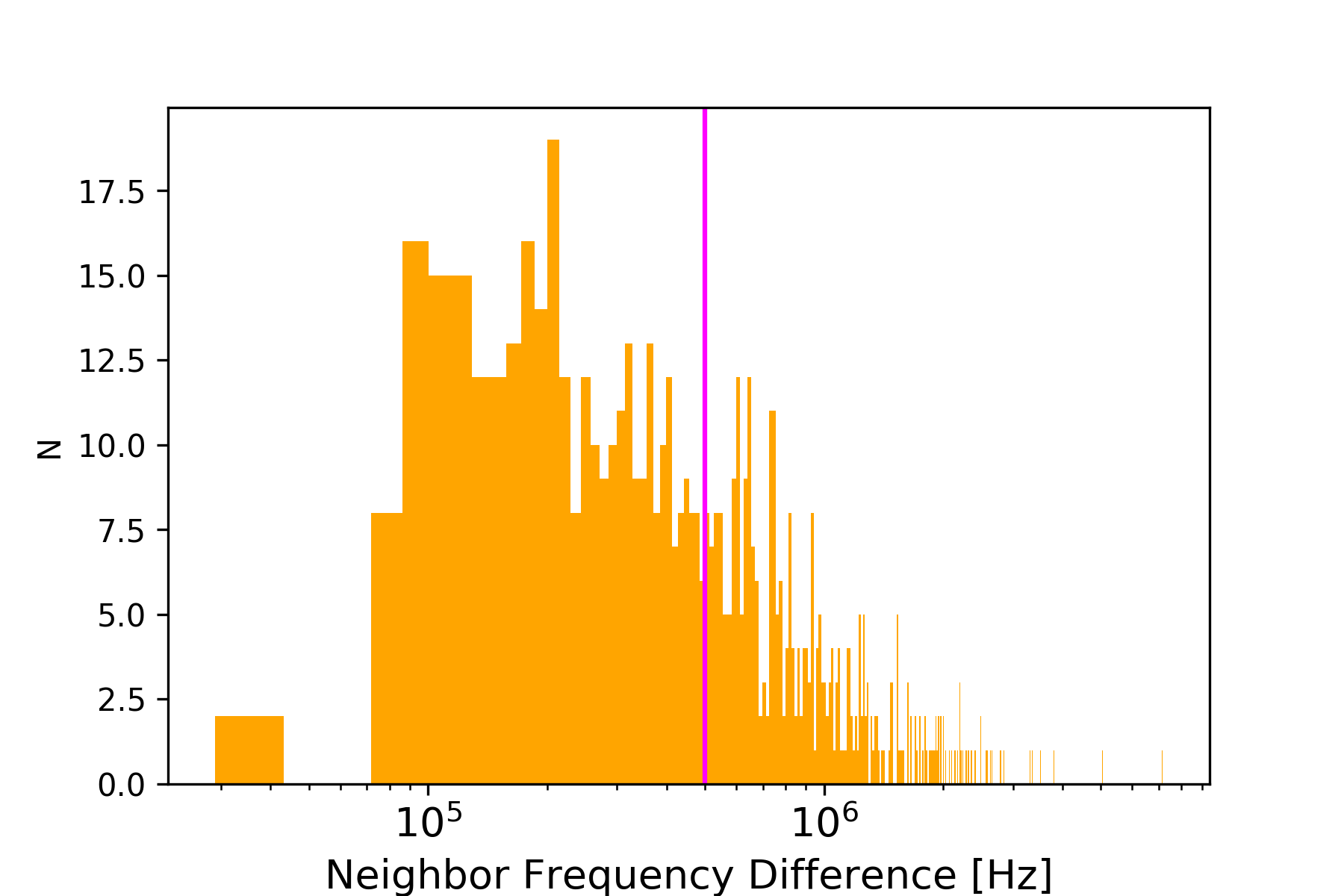}
\caption[Neighbor Freq Difference Histogram]{Neighboring resonator frequency difference from the BLAST-TNG 350$\mu$m array. The magenta vertical bar is at 500kHz which corresponds to the FFT bin width. Only $50.4\%$ of the found resonators are above this line; about half of the resonators share an FFT bin.}
\label{fig:neigh_freq_diff}
\end{figure}

The current frequency scatter for resonators with a 500MHz band centered at 750MHz assures that many will fall within the same bin ($\Delta f_{bin} = 500$kHz). A histogram of the difference between neighboring resonators from BLAST-TNG's 350$\mu$m array is shown in Fig. \ref{fig:neigh_freq_diff}. The vertical magenta line is at 500kHz which corresponds to the FFT bin width and only 50.4$\%$ of the found resonators are above this line. Thus we require buffering the FFT bins to select the same bin multiple times. Post-FFT fine channelization can then be used to recover each resonator to maintain high detector yield. Future efforts to utilize post fabrication capacitor editing have the potential to eliminate the need for this module. 

\subsubsection{Fine Channelization}
\label{sec:fine_channelization}
After the coarse channelization and bin selection we perform a fine channelization in two steps: digital down conversion, and vector accumulation. The first step is a time-division multiplexed complex multiply of an FFT bin value with a pre-calculated waveform. This waveform is the complex conjugate of the beat frequency between a tone and its corresponding FFT bin center frequency,
\begin{equation}
(f_{beat})^* = e^{-2\pi i(f_{tone}-f_{bin})t}.
\end{equation}
After this down conversion, the complex signal no longer oscillates due to this bin center offset and can now be coherently accumulated. If this step was not taken the signed values would accumulate quickly to zero.

The signals are then fed into a vector accumulator which accomplishes both an averaging and downsampling to reduce the output detector bandwidth. The number of accumulations can be programmed on the fly to allow for dynamic detector bandwidth sampling. 

\subsubsection{UDP packetizer}
The integrated detector data is collected into a dual port memory which safely crosses a clock domain and then is passed to a UDP packetizer. User Datagram Protocol (UDP) is a lighter Ethernet packet protocol that does not handshake like TCP. This allows for reduced gateware complexity and high bandwidth Ethernet streams. An open-source VHDL implementation of a UDP streaming Media Access Controller (MAC) from Mike Field is used\footnote{\url{https://github.com/hamsternz/FPGA_GigabitTx}}. The baseline design will use a four port 1GbE FMC card by Opsero/FMCEthernet\footnote{\url{https://opsero.com/product/ethernet-fmc/}} to simultaneously stream all four independent channels.

\subsection{uPC Software and Network}
\begin{figure}[!t]
\centering
\includegraphics[width=4in]{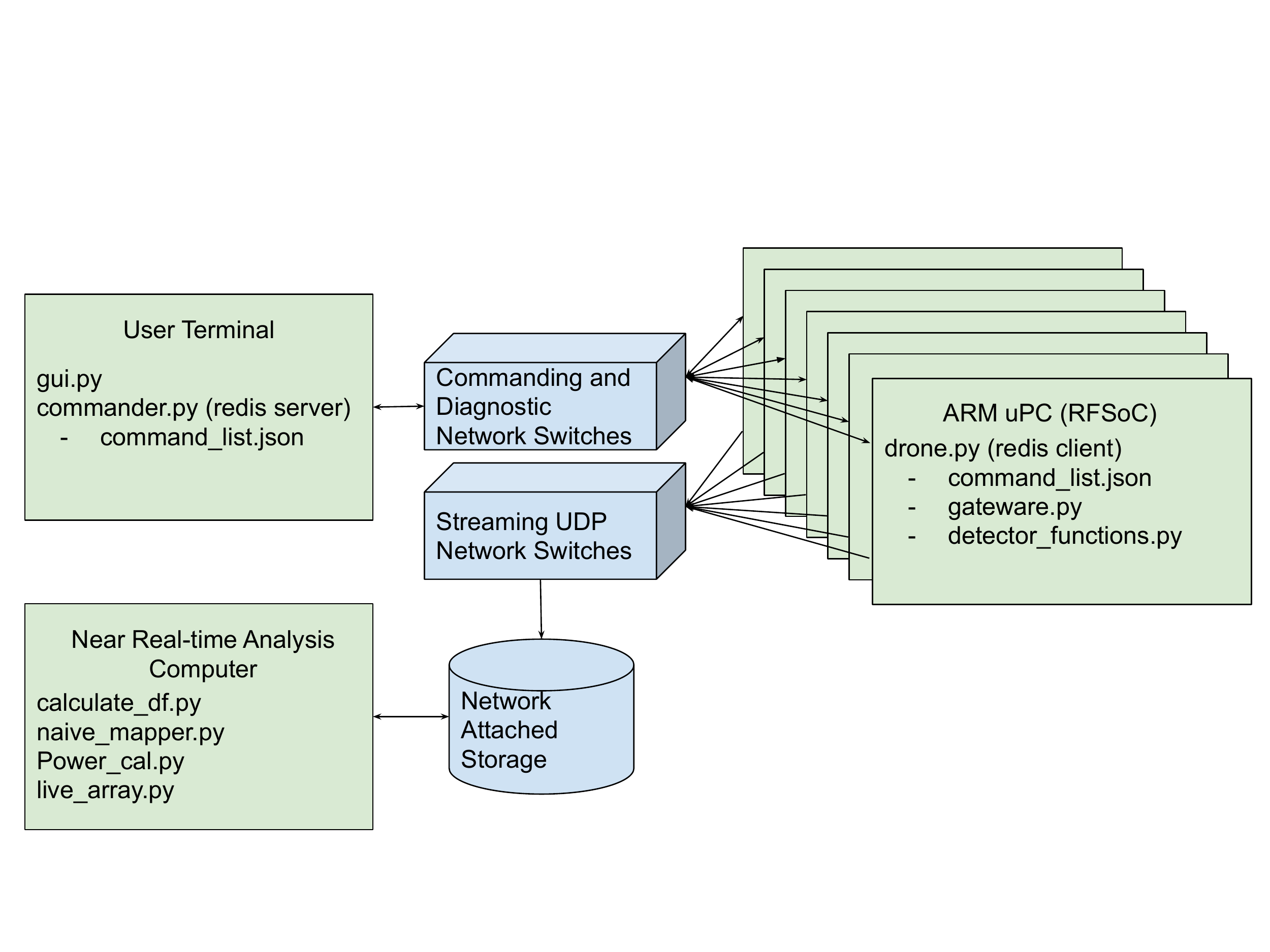}
\caption[Software Architecture]{A block diagram of the software and network architecture for the CCAT-prime instrument. The software uses a server-client model where the user terminal acts as the server and each readout ARM uP is a client. The user command terminal is physically connected to each RFSoC via an Ethernet network. The commanding Ethernet network is separated from the streaming UDP packet network. The UDP data is saved to a network attached storage device which is accessible for near real-time analysis.}
\label{fig:software_arch}
\end{figure}

A minimal implementation of the software and network architecture is shown in Fig. \ref{fig:software_arch}. 
The software and network must be discussed together as the commanding, diagnostics, and data capture are all occurring on different physical systems connected via an Ethernet network. Each RFSoC contains an ARM Cortex processor and will be configured to run the PYNQ image v2.6. The PYNQ image contains a Linux filesystem and a Python API to easily interact with the FPGA gateware and other hardware peripherals. Differing from the ROACH2 systems deployed in various instruments, the CCAT-prime readout takes advantage of the integrated microprocessors of each RFSoC. Each uPC acts as an independent computing node performing the majority of software computation and is provided with tasks from a command server, a model analogous to a distributed computing network. This method scales easily and eliminates the requirement of a high performance user terminal to perform the numerous computation tasks simultaneously. It also cuts down on the required data rate between the user terminal and each RFSoC. In normal operation only key value pairs will be needed to send command strings and status bits. Redis is an in-memory database and messaging service that uses TCP to send and receive messages in a server-client model. While written in C, Redis has Python bindings that are easy to integrate with the PYNQ framework. The built in messaging queue capabilities to publish and subscribe to various channels is ideal for commanding multiple RFSoCs over the network. The implementation is discussed in detail below.

\subsubsection{commander.py \& drone.py}
The two main Python programs implemented to control the readout are commander.py and drone.py. Commander.py runs on the user terminal and publishes commands to Redis, while drone.py runs as a Redis client and subscribes to the command channel and executes the received commands. This requires a command list to map the received key-value pair to a command.

Each RF network on the array will be controlled by an independent instance of drone.py. Each drone is given a unique designation that is linked to a DSP channel (1-4 in baseline) and RFSoC number in the server rack, these will eventually be mapped to detector array networks. The main functionality of drone.py is to execute commands received from subscribed Redis channels and publish status and exception flags. The required functions can be split broadly into two main files gateware.py and detector$\_$functions.py. The first is dedicated to the calls that are directly interfacing with gateware on the FPGA for example reading and writing to registers with custom drivers utilizing the PYNQ library. The detector functions would contain most of the functionality specific to kinetic inductance detectors, such as creating frequency combs, find and fit resonances from sweeps, and calculate $\Delta f$.

\subsection{Analysis Software}
Preliminary analysis software which provides useful data products and quick looks will be required for pre-deployment tests and commissioning. The first of which is to convert the raw in-phase and quadrature (I$\&$Q) data streams to something proportional to power. 

The shift in frequency is closely related to the power absorbed and is thus typically the metric to estimate. There are a few different methods for doing this and variations of each method: The IQ method which fits the resonance in the complex plane and moves its center to the origin and phase rotates such that the frequency response ends up in one quadrature. The change in signal can then be scaled by the derivatives of the loop with respect to frequency to convert the timestreams into a frequency shift. Another method finds the frequency shift assuming small perturbations from resonance which involves only taking derivatives of the complex sweep with respect to frequency, no fitting required, this is derived in \citenum{Barry2014}. Typical implementations of these methods are linear approximations to the frequency shift and thus offer only limited range of shifts up to $\sim$ 0.65 of a linewidth before accumulating an error of greater than 10\%. An equation which computes the fractional frequency shift which follows the non-linear resonator response is,
\begin{equation}
    x(t) = -\frac{Q(t)sin\theta_0}{2 Q_r (I(t)cos\theta_0 - A)},
\end{equation}
where the un-modulated phase on resonance is $\theta_0=arctan(Q/I)|_{f_0}$, $Q_r$ is the total quality factor, and A is resonator depth. This equation is found by inverting the resonator forward transmission Eq. \ref{eq:s21}(see appendix \ref{sec:app_inverse_method}). All three approaches will be explored in a future study.

A highly desirable goal of the time-domain science is to produce maps with a cadence of at least one day. While the maps do not need to be fully processed the important transient information must be easily detectable. The initial deployment data rate will be around 1 Gb/s which provides a challenging downstream processing task. A 10 GbE connected computer with a carefully planned memory access strategy and processing threads will serve as the analysis and storage computer. We will aim for an approach similar to \citenum{MSmith2016} for nearly real-time processing to produce quick maps nightly.

\section{Implementation and Resource Utilization}
\label{sec:utils}
\begin{figure}[!t]
\centering
\includegraphics[width=2in]{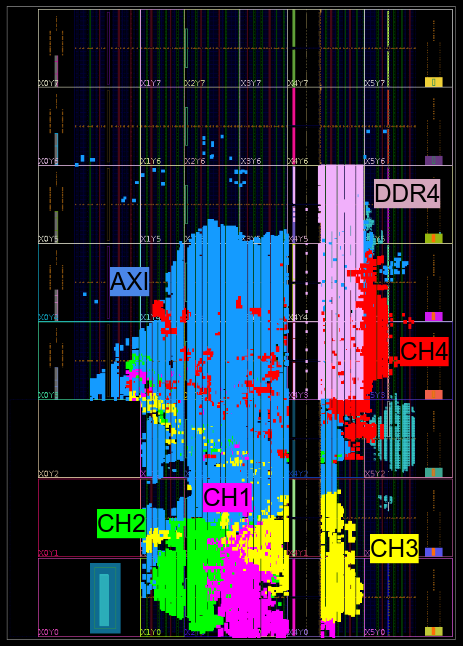}
\caption[Resource Utilization Device View]{Resource utilization of the baseline design in device logic cell view. Implemented four x 1000 channel design on the Zynq UltraScale+ ZCU28DR - with the four independent channels highlighted along with the DDR4 memory and AXI interconnects.
 }
\label{fig:resource_util_device_view}
\end{figure}

An important consideration when developing any digital design for an FPGA is the resource utilization in comparison to the resources available. The four channel baseline design was synthesised and implemented on the gen 1 RFSoC providing the following utilization report: CLB LUTs 22$\%$, CLB Regs 18$\%$, DSPs  6$\%$, Block RAM 17$\%$, URAM 10 $\%$. This is visually represented in Fig. \ref{fig:resource_util_device_view} in a logic fabric view of the chip. Each channel is highlighted along with the AXI interconnects and DDR4 memory bank. Two things are immediately apparent from this visual representation: the AXI interconnects are taking a significant area of the fabric, and there is appears to be ample room to grow in the north west region of the fabric. The utilization report numbers support the visual assessment and provide encouraging proof that a doubling of the instantaneous bandwidth from 500 MHz to 1 GHz should be possible.

\begin{figure}[!t]
\centering
\includegraphics[width=2in]{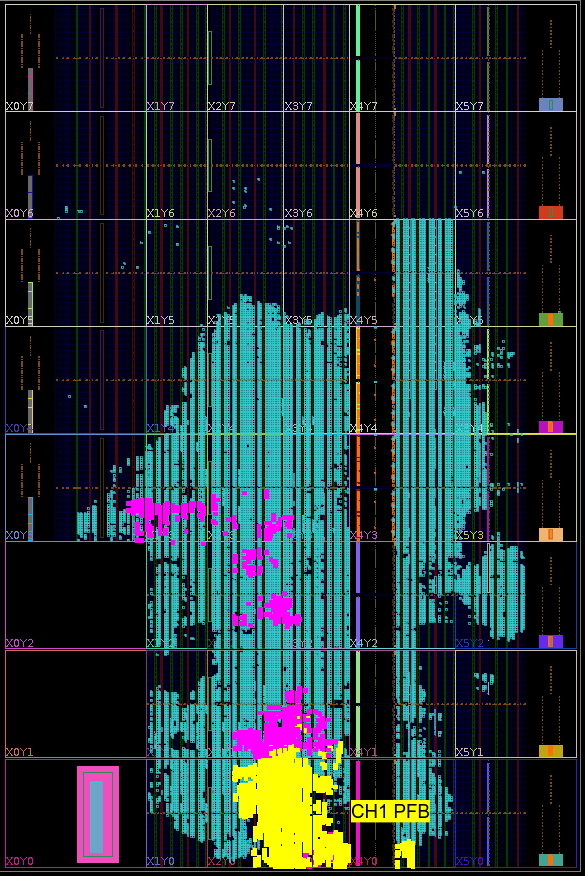}
\caption[Channel 1 PFB utilization]{Device view of the resource utilization with only channel 1 highlighted in magenta and yellow. The yellow highlights the resources used solely for the polyphase filterbank (PFB). The remaining magenta elements are the rest of the resources used for channel 1. Clearly the fabric area covered by the PFB dominates the utilization of the channel.
 }
\label{fig:chan1_pfb_util_device_view}
\end{figure}

Upon closer examination into the resource utilization of a single channel we see that the PFB dominates the fabric area. Figure \ref{fig:chan1_pfb_util_device_view} gives the device fabric view of the four channel design with only a single channel highlighted in magenta and yellow. The yellow highlights solely the PFB resources and the magenta highlights the remaining parts of the channel. The utilization report of the PFB gives: 4303 CLB LUTs, 9844 CLB Registers, 675 CARRY8, 1407 CLB, 3375 LUTs as Logic, 928 LUTs as Memory, 27.5 Block RAM Tiles, and 52 DSPs. From the analysis of the digital crosstalk requirements described earlier in Sec. \ref{sec:xtalk_reqs}, the front end filterbank may not be necessary given the high number of accumulations in the fine channelization and expected resonator crosstalk level. This will allow for larger FFTs within the resource budget.

\section{Demonstrations}
\begin{figure}[!t]
\centering
\includegraphics[width=3in]{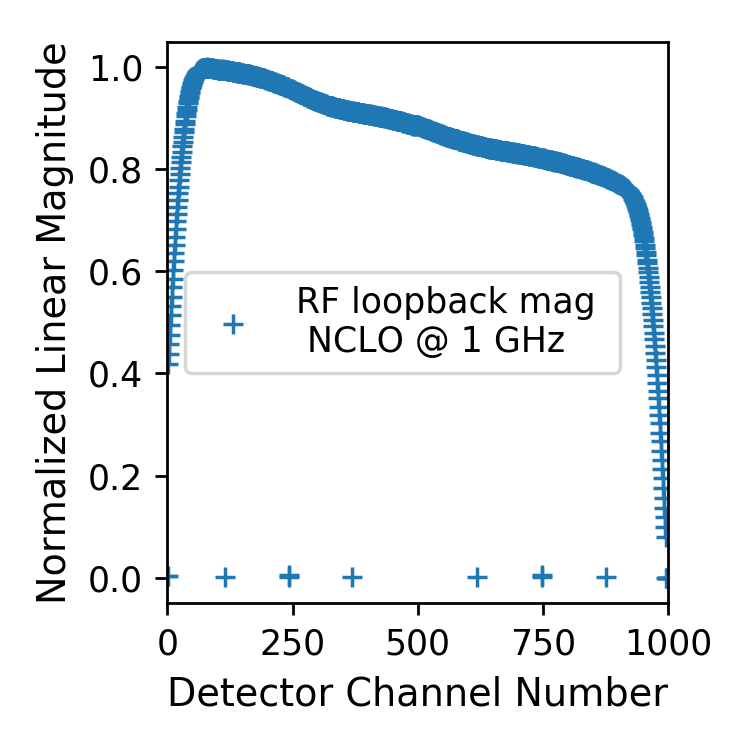}
\caption[VNA comb loopback]{RF loopback measurement with the numerically controlled local oscillator (NCLO) set to 1GHz and a 1000 tone comb.}
\label{fig:loopback_mags_vna_comb}
\end{figure}

The baseline design was set up in RF loopback with an SMA cable connecting a D/A to it's corresponding A/D. One thousand tones spanning -250 to 250 MHz at baseband (pre-mixer) were generated and up-converted to 750 - 1250 MHz with the numerically controlled local oscillator (NCLO) within the digital mixer set to 1 GHz. On the receive side after being sampled by the A/D the waveform is down-converted to baseband by another NCLO set to 1 GHz. Each tone is assumed to be an independent detector channel and is demodulated and accumulated within the digital signal processing chain. The output of the last stage of the signal processing is captured and displayed in Fig. \ref{fig:loopback_mags_vna_comb}. All one thousand channels are shown and the plot is normalized to the maximum. A frequency dependent slope is seen in the magnitudes and a roll-off on both ends of the band. The roll-off is suspected to arise from the half-band up and down sampling filters frequency response within the RF data converters. Digital leveling can be achieved by applying the inverse transfer function to the amplitudes of the tones in the arbitrary waveform.

\begin{figure}[!t]
\centering
\includegraphics[width=4.5in]{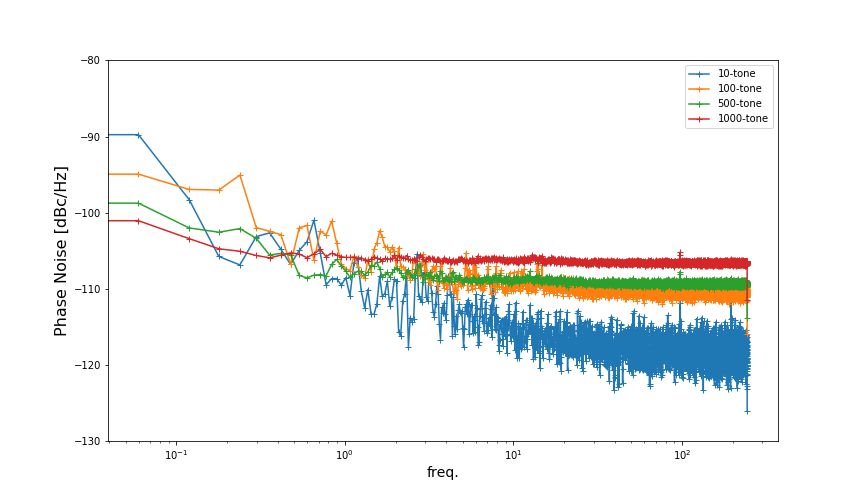}
\caption[]{RF loopback phase noise measurement for various numbers of tones with a common mode removed.}
\label{fig:multi_tone_phase_noise}
\end{figure}

To characterize the dynamic range and phase noise a loopback measurement was made with a single chain of the design. The I\&Q for each detector channel are streamed in UDP packets and captured for two minutes with a sample rate of $\sim500$ Hz. The phase noise power spectral density is then calculated on the time-stream either by first calculating the phase time-stream with $\phi = arctan(Q/I)$ then finding the power spectral density, or by first calculating the power spectral density of I\&Q separately ($S_{II}$ and $S_{QQ}$) then using the following equation,
\begin{equation}
 S_{\phi \phi} = \frac{1}{2}\frac{S_{II} + S_{QQ}}{ \overline{I}^2 + \overline{Q}^2},
\end{equation}
Where the sum of the power spectral densities are normalized by the mean magnitude squared. Figure \ref{fig:multi_tone_phase_noise} shows the loopback phase noise power spectral density in dBc/Hz for various numbers of tones in the arbitrary waveform and presents the median detector channel from each. A common mode present across each channel was removed with a single template produced with a principle component analysis. The inverse of the phase noise power spectral density is the maximum possible signal to noise or dynamic range achievable for that channel and specific arbitrary waveform. For 1000 tones this system will be detector noise limited for shifts of up to $\sim 3/4$ of a linewidth based on Fig.$~$\ref{fig:snr_vs_frac_freq}. 

\section{Conclusion}
The RFSoC based readout for the Mod-Cam and Prime-Cam instruments was presented. The requirements were derived from expected detector parameters. The hardware, gateware, and software of the baseline design were described in detail. The implementation and resource utilization for the design proved that 4x 1000 channel design fit comfortably within the device fabric. This also revealed that there is space to increase the bandwidth for the 850 GHz instrument. We demonstrated the operation of the design in RF loopback showing that the digital mixers can replace analog IQ mixers and that the phase noise is low enough with 1000 tones to ensure detector noise limited performance for the anticipated operational range. In the near term, we expect to have an 850 GHz prototype array in our test cryostat at NRC Herzberg to inform the optimization of all aspects of the RFSoC based system. The first field deployment of the RFSoC readout will be with the Mod-Cam instrument in 2024. 

\appendix    

\section{Derivations}
\label{sec:appendix}

\subsection{Equivalent Output Noise Temperature}
\label{sec:eq_det_noise}
Starting with the equation for the complex transmission of an LRC resonator \cite{Mauskopf2018},
\begin{equation}
    S_{21}(f) = 1 - \frac{Q_r}{Q_c}\frac{1}{1+j2Q_r(f-f_0)/f_0},
\end{equation}
Where $f$ is frequency, $f_0$ is the resonant frequency, $Q_r$ is the total quality factor, $Q_c$ is the coupling quality factor.
Rewriting the fractional frequency as $(f-f_0)/f_0 = x$ and splitting the above equation into real and imaginary components we get,
\begin{equation}
Re\{ S_{21}\} = 1 - \frac{Q_r}{Q_c (1 + 4 Q_r^{2} x^2)},
\end{equation}
and
\begin{equation}
Im\{ S_{21}\} = -\frac{2 Q_r^{2} x}{Q_c (1 + 4 Q_r^{2}x^2 )}.
\end{equation}
Taking the derivative of each with respect to fractional frequency $x$, 
\begin{equation}
\frac{d Re\{ S_{21}\}}{dx} = \frac{8 Q_r^{3} x}{Q_c (1 + 4 Q_r^{2} x^2)^{2}},
\label{eq:dRef}
\end{equation}
\begin{equation}
\frac{d Im\{ S_{21}\}}{dx} = \frac{2 Q_r^{2}}{Q_c (1 + 4 Q_r^{2} x^2)} - \frac{16 Q_r^{4}x^2}{Q_c (1 + 4 Q_r^{2} x^2)^{2}}.
\label{eq:dImf}
\end{equation}
In typical operation the measured output voltage depends on the probe tone input voltage and the forward transmission, 
\begin{equation}
V_{out} = V_{in}S_{21}(x).
\end{equation}

The absorption of optical photons will shift the resonant frequency modulating the transmission. Taking the derivative of the above equation to find the change in output voltage with respect to absorbed optical power or voltage responsivity,  
\begin{equation}
\frac{dV_{out}}{dP} = V_{in}\frac{dS_{21}}{dP} = V_{in} \frac{dS_{21}}{dx}\frac{dx}{dP}.
\end{equation}
Expanding the complex terms,
\begin{equation}
\frac{dV_{out}}{dP} = V_{in}\frac{dx}{dP}\Big( \frac{d Re\{ S_{21}\}}{dx}  + j\frac{d Im\{ S_{21}\}}{dx} \Big).
\end{equation}
Following \cite{Sipola2019} we can then calculate the noise temperature as,
\begin{equation}
    T = \frac{(R \times NEP)^2}{k Z_0},
\end{equation}
where NEP is the detectors noise equivalent power, k is boltzmann's constant, $Z_0$ is the characteristic impedance, and $R = dV_{out}/dP$ is the voltage responsivity.

\subsection{Fractional Frequency Shift by Inverse Method}
\label{sec:app_inverse_method}
The equation for the complex forward transmission of a shunt resonator is,
\begin{equation}
S_{21} = 1 - \frac{Q_r}{Q_c} \frac{1}{1 + j 2 Q_r (f-f_0)/f_0},
\label{eq:s21}
\end{equation}
Where $Q_r$ is the total quality factor, $Q_c$ is the coupling quality factor, $f_0$ is the unmodulated resonant frequency. The term $(f-f_0)/f_0$ is the fractional frequency shift commonly written as $x$.
The fractional frequency shift $x$ can be directly related to the power absorbed through measurements of the responsivity.
Fixed tone readout systems estimate this value by measuring the instantaneous complex forward transmission at the unmodulated resonant frequency $f_0$. These complex values are named the in-phase $I$ and quadrature $Q$ components of the transmission,
\begin{equation}
S_{21}(f_0) = Re\{S_{21}(f_0)\} + j Im\{S_{21}(f_0)\} = I(t) + j Q(t).
\label{eq:s21toIQ}
\end{equation}
We seek an equation which gives the fractional frequency $x$ with the measured I/Q signals as an input. This requires inverting Eq. \ref{eq:s21} and using above relation to I/Q.

Equation \ref{eq:s21} can be split into real and imaginary components as,
\begin{equation}
S_{21}= 1 - \frac{Q_r}{Q_c (1 + 4 Q_r^2 x^2)} + j \frac{2 Q_r^{2} x }{Q_c (1 + 4 Q_r^2 x^2)}.
\end{equation}
This makes it easier to see that subtracting 1 from the real term and dividing by the imaginary term simplifies the equation to,
\begin{equation}
    \frac{Re\{S_{21}\}-1}{Im\{S_{21}\}} = - \frac{1}{2 Q_r x}.
\end{equation}
Now using the relation from Eq. \ref{eq:s21toIQ} and solving for x,
\begin{equation}
    x = -\frac{Q(t)}{2Q_r(I(t) - 1)}.
\end{equation}
This equation assumes that a shift in resonant frequency is equivalent to a shift of stimulus tone frequency in the opposite direction. It's also an idealized symmetric resonator which has been normalized and phase rotated such that the imaginary component on resonance is zero.

In practice the resonator measurements have some arbitrary magnitude scaling and phase rotation angle. Including these effects the fractional frequency shift equation becomes,
\begin{equation}
    x = -\frac{Q(t)sin\theta_0}{2Q_r(I(t)cos\theta_0 - A)}.
\end{equation}
Where A is the resonator depth, $\theta_0$ is the phase angle at the resonant frequency $f_0$ found from $\theta_0 = arctan(Q/I)$. 

\acknowledgments %
The CCAT-prime project, FYST and Prime-Cam instrument have been supported by generous contributions from the Fred M. Young, Jr. Charitable Trust, Cornell University, and the Canada Foundation for Innovation and the Provinces of Ontario, Alberta, and British Columbia. The construction of the FYST telescope was supported by the Gro{\ss}ger{\"a}te-Programm of the German Science Foundation (Deutsche Forschungsgemeinschaft, DFG) under grant INST 216/733-1 FUGG, as well as funding from Universit{\"a}t zu K{\"o}ln, Universit{\"a}t Bonn and the Max Planck Institut f{\"u}r Astrophysik, Garching.
Prime-Cam:
The construction of EoR-Spec is supported by NSF grant AST-2009767. The construction of the 350 GHz instrument module for Prime-Cam is supported by NSF grant AST-2117631.
Chai:
The CHAI instrument is supported by DFG grant CRC 956/3, project ID 184018867 as well as funding from Universit{\"a}t zu K{\"o}ln. 

The author would also like to thank Sam Rowe, Jack Hickish, Jenny Smith, Jeb Bailey III, Leandro Stefanazzi, and Jozsef (Mazsi) Imrek for collaborative readout discussions, Ed Chapin, Lewis Knee, Doug Henke, Frank Jiang, and Doug Johnstone of the Herzberg Astrophysics and Astronomy Research Centre for use of the mm-wave lab equipment and sage advise.

\bibliography{main} 
\bibliographystyle{spiebib} 

\end{document}